\UseRawInputEncoding 
\documentclass[pra,onecolumn,superscriptaddress]{revtex4}%
\usepackage{amssymb}
\usepackage{amsmath}
\usepackage{graphicx}
\usepackage{dcolumn}
\usepackage{xcolor}
\usepackage{bm}
\usepackage{subfigure}
\usepackage{amsfonts}
\usepackage{appendix}
\usepackage{tikz}
\usetikzlibrary{quantikz2,backgrounds,fit,decorations.pathreplacing}
\usepackage{xcolor}%
\providecommand{\U}[1]{\protect\rule{.1in}{.1in}}

\begin{document}
\preprint{APS/123-QED}

\title{Enhancing the performance of Variational Quantum Classifiers with hybrid autoencoders}

\author{Georgios Maragkopoulos}

\affiliation{Department of Informatics and Telecommunications, National and Kapodistrian
University of Athens, Panepistimiopolis, Ilisia, 15784, Greece}

\author{ Aikaterini Mandilara}

\affiliation{Department of Informatics and Telecommunications, National and Kapodistrian
University of Athens, Panepistimiopolis, Ilisia, 15784, Greece}
\affiliation{Eulambia Advanced Technologies, Agiou Ioannou 24, Building Complex C, Ag. Paraskevi, 15342, Greece}
\author{ Antonia Tsili}

\affiliation{Department of Informatics and Telecommunications, National and Kapodistrian
University of Athens, Panepistimiopolis, Ilisia, 15784, Greece}

\author{ Dimitris Syvridis}

\affiliation{Department of Informatics and Telecommunications, National and Kapodistrian
University of Athens, Panepistimiopolis, Ilisia, 15784, Greece}
\affiliation{Eulambia Advanced Technologies, Agiou Ioannou 24, Building Complex C, Ag. Paraskevi, 15342, Greece}

\begin{abstract}

Variational Quantum Circuits (VQC) lie at the forefront of quantum machine learning research. Still, the use  of quantum networks for real data processing remains  challenging  as the  number of available qubits cannot accommodate a large dimensionality of data  --if the usual angle encoding scenario is used. To achieve dimensionality reduction, Principal Component Analysis  is routinely  applied as a pre-processing method  before the embedding of the classical features on qubits. In this work, we  propose an alternative method which reduces the dimensionality of a given dataset by taking into account the specific quantum embedding that comes after. This method aspires to make quantum machine learning with VQCs more versatile and effective on   datasets of high dimension. At a second step,  we propose a quantum inspired classical autoencoder model which can be used to encode information in low latent spaces. The power of our proposed models is exhibited via numerical tests. We show that our targeted dimensionality reduction method considerably  boosts  VQC's performance and we also identify cases for which the second model outperforms classical linear autoencoders in terms of reconstruction loss.

\end{abstract}
\maketitle

\section{Introduction}

Quantum hardware  reaches new milestones faster than expected and IBM's roadmap indicates that quantum computers will be  capable of running $1$ billion gates beyond $2033$ \cite{QAE5}.
Still, Quantum Machine Learning (QML) faces numerous challenges not only at hardware level but also at design level,  with the domain  trying to compete with, the  more established  in real world applications, classical approaches \cite{QAE6}. One of the challenges  concerns the encoding of high-dimensional classical data into qubit states. In literature one mainly encounters four  types of encodings: basis, amplitude, Hamiltonian and angle encoding \cite{QAE18} --also known as time-evolution or rotation encoding.  While amplitude encoding is the most efficient in terms of qubit resources, popular packages \cite{QAE19, QAE20} reveal that at this moment, angle encoding is the most widely used for QML models built with Variational Quantum Circuits (VQCs). 
However, considering angle encoding, the number of qubits is, in practice, as large as the dimension of the embedded data \cite{QAE4}, making the task intractable both for simulations and experiments.
One way to reduce the number of qubits is to increase the number of features per qubit by performing
repetitions of layers  across the quantum circuit \cite{QAE1}. In  a second approach  Principal Component Analysis (PCA) method is used to reduce the dimensionality before embedding  the extracted components to  a VQC  \cite{QAE2,QAE28}. Still the question  whether more adapted methods exist for reducing the dimension of classical data before quantum embedding remains. 

To address this question, different suggestions have been put forward and tested during the last years. In \cite{QAE12} pre-trained models are used as a preface for quantum circuits, enhancing the ability of the last to classify images. In \cite{QAE15}, a supervised hybrid system  optimizes quantum embedding with the help of classical deep learning. In \cite{QAE9}, the authors integrate classical vision transformers in a VQC to create hybrid units which are then jointly trained. This hybrid model is a supervised learning model, demonstrating promising results. 
In recent works \cite{QAE39,QAE40}, explore different flavours of angle encoding in various datasets, providing further evidence on angle encoding being the most prevalent encoding method. 

In this work, we propose a new method  for enhancing the encoding of high dimensional classical data into qubits by adjusting the idea of classical autoencoders (AE) to quantum variational methods, presented in Section ~\ref{S3}. We call the model  \textit{PCA-embedded quantum autoencoder} (PQAE), since this consists from a classical autoencoding structure, the quantum data-embedding block of the VQC 
under study and a kernel PCA (KPCA) method that uses the generated quantum kernels. 
In contrast to existing encoding methods for QML, the proposed PQAE works with datasets of arbitrary size and complexity, by creating an autoencoder which re-generates the dataset with  a bottleneck layer that has the same structure as  the data embedding scheme of the VQC. A classical KPCA method is interjected after the bottleneck  enabling this hybrid method to treat a big number of qubits.  
In addition, we propose a quantum inspired classical model for dimesionality reduction as a secondary model, named quantum-inspired autoencoder (QAE). 
QAEs shares many similarities with PQAEs, yet their advantages come from a different scope. QAEs are classical models which regenerate the dataset using a quantum inspired transformation in the `bottleneck' layer which changes the data representation. 

The numerical results of Section~\ref{S4} show that PQAEs  can give an important boost in the VQCs performance as compared to PCA method. QAE also does achieve lower reconstruction error than common linear AEs in four typical classification tasks: Iris, Wines, Seed and (binary) MNIST datasets. Let us finally note that both proposed methods are unsupervised learners, meaning that the encoding is indifferent to the target of the task, thus the same encoded data could also be used for a range of classification tasks or regression.

Various methods in literature have claimed the name `quantum autoencoder', some of which we list in the following. In \cite{QAE10}, a quantum autoencoder is trained to compress quantum states, a task beyond classical capabilities. The compression occurs inside the quantum circuit, making it a quantum input to quantum output autoencoder, which differs from our classical input to classical output approach. In \cite{QAE11}, the authors directly follow the proposed scheme in \cite{QAE10}, and the quantum autoencoder is followed-up by a quantum classifier. This has a similar flavour to our approach, as they use a classifier after the encoder. In \cite{QAE8}, the authors construct a  fully quantum variational autoencoder, which is applicable for both classical and quantum data compression. Their proposed model is a direct translation of classical autoencoders in a quantum computer, in contrast, the models that we propose offer a hybrid approach that is not associated with an existing classical model.  In \cite{QAE13}, a model where a classical autoencoder is dressed with a quantum circuit is introduced. In this model, the trainable weights are present in both the classical and quantum parts of the model, in our approach we only train 
`classical weights' and use KPCA as well, with the goal of passing the data in the VQC. Finally,  quantum inspired activation functions are used in \cite{QAE14}, as the authors demonstrated that quantum activation functions are efficient in feature selection of input images, which showcase the potential of applying quantum inspired methods as we do with the QAE model.

\section{Basic Concepts and Methods\label{S2}}
In order to make this work self-contained, we briefly describe here the basic classical and quantum methods
which we use to build the new models.

\subsection{ PCA method and Classical Autoencoders}

PCA is a statistical technique used for dimensionality reduction, widely applied in data analysis and Machine Learning (ML) \cite{QAE25}. It aims to transform a dataset with a high-number of features into a lower-dimensional form while retaining as much variability  as possible. 
Given \( n \),   $N$-dimensional samples, $\mathbf{x}_i\in \mathbb{R}^{N}$ with $i=1, \ldots,n$,   one first forms the data matrix  \( \mathbf{X} \in \mathbb{R}^{n \times N}\).  PCA begins by computing the covariance matrix \( \mathbf{C} \in \mathbb{R}^{N \times N} \) of  \( \mathbf{X} \), given by:
\begin{equation}
\mathbf{C} = \frac{1}{n} \mathbf{X}^\top \mathbf{X}~.\label{eq1}
\end{equation}
 The eigenvectors \( \mathbf{v}_1, \mathbf{v}_2, \ldots, \mathbf{v}_N \in \mathbb{R}^{N}\) of \( \mathbf{C} \) capture the directions of maximum variance in the data while the corresponding eigenvalues  \( \lambda_1, \lambda_2, \ldots, \lambda_N \)
 are assumed in descending order. One chooses the first $k$ eigenvectors, with $k<N$ as the principal components and uses them 
to reduce the dimension of the initial data from $N$ to $k$ by projecting 
the data matrix \( \mathbf{X}\) on the space spanned by the $k$ $N$-dimensional principal components.

KPCA  \cite{QAE26} offers a flexible approach as compared to traditional PCA, when the relationships between the data variables  $\mathbf{x}_i$ are nonlinear. It operates by mapping the data into a higher-dimensional feature space using a kernel function \( k(\mathbf{x}_i, \mathbf{x}_j) \) that computes the similarity between the data points \( \mathbf{x}_i \) and \( \mathbf{x}_j \). In KPCA, instead of computing the covariance matrix directly, the kernel matrix \( \mathbf{K} \in \mathbb{R}^{n}\) with elements \( K_{ij} = k(\mathbf{x}_i, \mathbf{x}_j) \) is computed. The eigenvectors \( \mathbf{a}_1, \mathbf{a}_2, \ldots, \mathbf{a}_n \) and corresponding eigenvalues \( \alpha_1, \alpha_2, \ldots, \alpha_n \) of \( \mathbf{K} \) capture the principal components in the higher-dimensional space. To reduce dimensions using KPCA, one simply selects the eigenvectors of the kernel matrix with the highest eigenvalues.

\begin{figure*}[t]
    \centering
    \hspace*{-0.25\textwidth}
    \includegraphics[ width=1.1\textwidth, height=0.4\textheight]{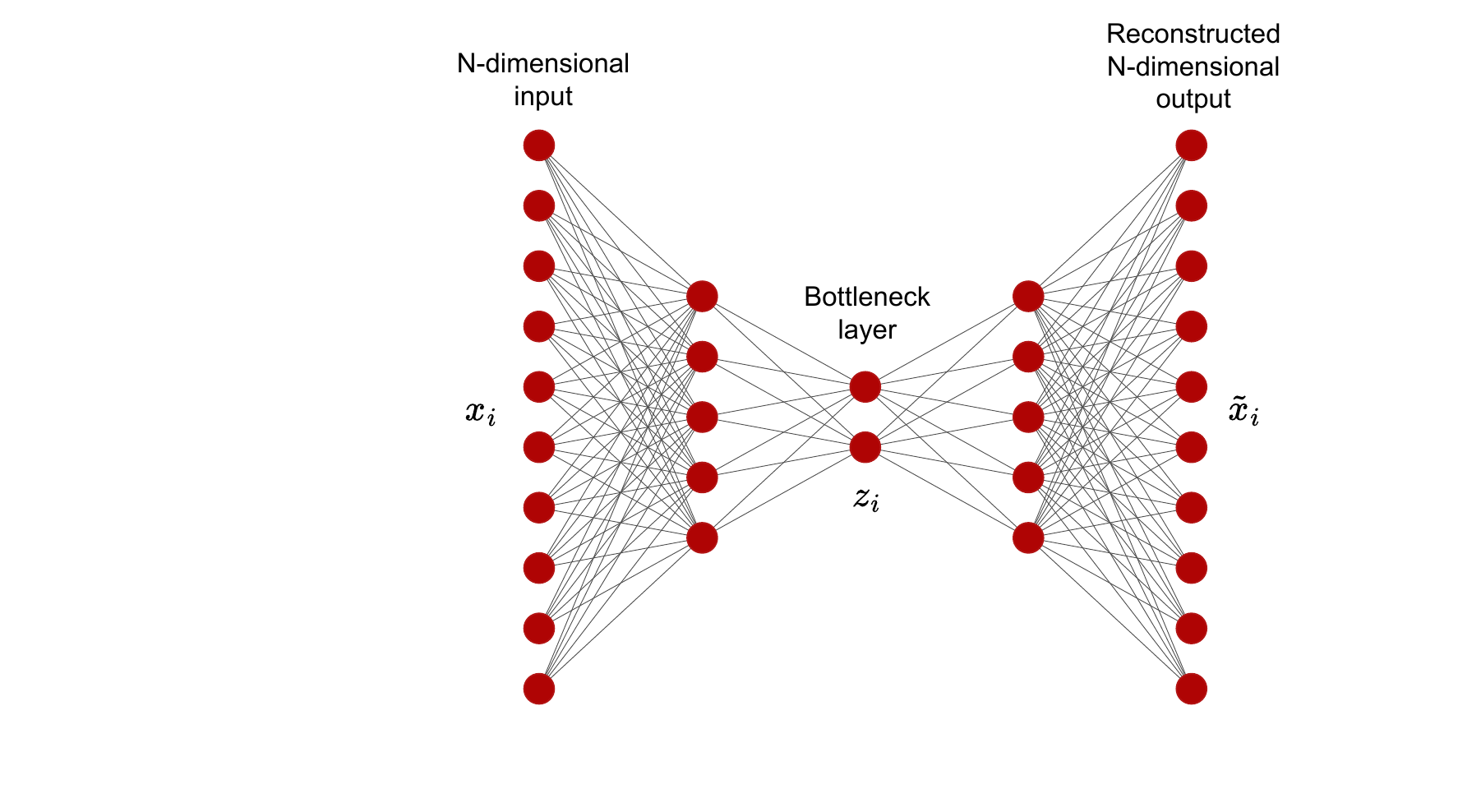}
     \caption{An AE with $L=2$ encoding and decoding layers. The $N$-dimensional  data point $\mathbf{x}_i$ is transformed  to a $k$-dimensional reduced representation $\mathbf{z}_i$  and then back to an $N$-dimensional  image $\mathbf{\Tilde{x}}_i$. The weights in the network are adjusted so that the reconstruction error between initial data and image data is minimum. }
    \label{fig1}
\end{figure*}

AEs are a type of Neural Network (NN) used for unsupervised learning, particularly employed in the field of representation learning and dimensionality reduction \cite{QAE24}. The basic function of an AE is to learn a compressed representation of the input data. This is achieved by first reducing the dimensionality of the input, then reconstructing the original input from this representation, and train the weights in the model so that  a loss function between input and output is minimized, see Fig.~\ref{fig1}.

In more detail, the encoder part of an AE network takes an input data $\mathbf{x}_i \in \mathbb{R}^{N}$ and maps it to a lower-dimensional latent space representation $\mathbf{z}_i \in \mathbb{R}^{k}$. This process involves a series of transformations, usually implemented through $L$ layers of NN units, represented as:
\begin{equation}
\mathbf{z}_i = E(\mathbf{x}_i) = E_{L}(E_{L-1}(\ldots E_{1}(\mathbf{x}_i) \ldots ))~.\label{eq2}
\end{equation} where $E_j$ with $j=1,\ldots,L$ denotes the $j$-th layer of the encoder.
The decoder part of the AE network takes the compressed representation $\mathbf{z}_i$ (output of the encoder) and attempts to reconstruct the original input data $\mathbf{x}_i$ from it. The reconstruction process involves a series of inverse transformations, typically implemented through layers that mirror the encoder, represented as:
\begin{equation} \mathbf{\Tilde{x}_i} = D(\mathbf{z}_i) = D_{0}(D_{1}(\ldots D_{L-1}(\mathbf{z}_i) \ldots )) \end{equation} where $D_j$ where $j=1,\ldots,L$ denotes the $j$-th layer of the decoder, and $\mathbf{\Tilde{x}_i}$ is the reconstructed output. The goal is to make $\mathbf{\Tilde{x}_i}$ as close to $\mathbf{x}_i$ as possible by minimizing a loss function. For the analysis in this work we  employ the loss function of Mean Squared Error (MSE) defined as
\begin{equation} \text{MSE} = \frac{1}{n} \sum_{i=1}^{n} (\mathbf{x}_i - \mathbf{\Tilde{x}}_i)^2~~. \end{equation}

AEs can be used as an alternative method to PCA for dimensionality reduction. Interestingly in \cite{QAE16,QAE17}  is proven that  PCA is equivalent to shallow AEs consisting of only linear activation functions.  

\subsection{VQCs and Quantum Kernels}

VQCs stand for quantum circuits built under specific architectural  ansatzes and with parametrized/variational quantum gates within.
 VQCs were introduced with the conception of Variational Quantum Algorithms (VQAs) \cite{QAE27} which are hybrid (quantum-classical)  algorithms  designed to address a big variety of computational  tasks by performing optimization in the parameters of a VQC.  A VQA adjusts the design of the    VQC, the measurements process on the output qubit states, the classical cost function built from these outputs and  the classical optimization techniques to the specific problem under study. 

In this work, we deal with the complex issue of encoding classical data to be processed in a VQC in the context of classification \cite{QAE18}, a supervised ML task. For such tasks, the simplest form of a VQC consists of two components:  the data-embedding or encoding block  and the variational or parametrised block. The data encoding block is the part of the quantum circuit where a classical data $\mathbf{x}$ is encoded into the qubits' state $|\phi(\mathbf{x}) \rangle$,  generating  the  quantum  feature map: $\mathbf{x}\rightarrow |\phi(\mathbf{x}) \rangle $ from the real space to the Hilbert space. In the variational block a sequence of parameterized quantum gates is applied on the encoded quantum state, and then the parameters, $\vec{\theta}$ of these gates are optimized through a classical optimization process. The unitary action of the encoding block on the quantum states, followed by the variational part of the circuit can be described as $|\psi_{out} \rangle = \hat{V}(\vec{\theta}) \hat{U}(\phi(\mathbf{x})) |\psi_{in} \rangle$ where habitually $|\psi_{in} \rangle=|00\ldots0 \rangle$. By repeated measurements on    the quantum state  $|\psi_{out}\rangle$ for a selected observable,  one can derive probabilities for each label as for an NN.

Since in our model we mostly occupy with the encoding block of the VQC, let us focus on the encoding methods. Generally speaking, the most ``qubit'' efficient encoding method is the amplitude encoding method where a $2^n$ dimensional normalized vector $\mathbf{x}$ is mapped on the state of $n$ qubits as $|\phi(\mathbf{x})\rangle = \sum_{i=1}^{2^n} x_i |i\rangle$. However while there are algorithmic procedures for preparing the quantum state $|\phi(\mathbf{x})\rangle$ the number of gates in the related circuits is increasing exponentially with $n$ \cite{QAE21,QAE22}. In this work, we assume angle encoding in VQCs as this is the most `physical' encoding method, using classical features to parametrise angles of rotation gates which are in turn experimentally tunable parameters.
More specifically, we implement  Pauli Feature Maps, meaning that the classical data are embedded in the quantum circuit via angle encoding in single-qubit gates using the Pauli operators $\hat{\sigma}_x$, $\hat{\sigma}_y$, $\hat{\sigma}_z$. For example, the trivial case of this encoding method leverages the product of a feature $x \in \left[0,1\right]$ with the Pauli operator  $\hat{\sigma}_y$  as $R_y(x \pi)=\mathrm{e}^{-i(x \pi\hat{\sigma}_y/2) }$. This expresses the incorporation of the feature $x$ in  the  gate $R_y(\phi)$ that  rotates the Bloch vector of a state around the y axis for an angle $\phi$. Such angle encodings can be performed using multiple features and Pauli gates, as $e^{-i\sum{x_i\hat{\sigma}_i}}$, increasing the complexity of the feature map \cite{QAE3}.
In addition to single qubit gates  used for the embedding of data,  the encoding circuit also includes entangling gates for exploiting in full extend the Hilbert space  and further increasing the complexity of the quantum feature map.

Finally, another flavour of KPCA can emerge if  classical kernels are replaced by quantum kernels. The latter are expressed as
\begin{equation}
k(x, x') = |\langle \phi(x) | \phi(x') \rangle|^2
\end{equation}
and can be understood as a similarity measure between data points $\mathbf{x}$, $\mathbf{x'}$  in the  space which emerges from the quantum feature map of an embedding circuit: $\mathbf{x}\rightarrow | \phi(\mathbf{x}) \rangle$, \cite{QAE18}.
The elements of a quantum kernel matrix can be experimentally estimated
via the  SWAP or  Inversion tests and then have the same use as classical ones \cite{QAE29,QAE4}. Quantum kernels become particularly interesting when 
 the quantum feature map is not classically computable  and in consequence  quantum kernels are speculated to provide  quantum advantage \cite{QAE30,QAE31,QAE32}.  
In this work we employ KPCA with quantum kernels for another aim. We employ them as a method of extracting  information from  the data-embedding part of the VQC  and  the input data. In addition, under the assumption that there is access to the quantum circuit and possibility for performing SWAP tests KPCA with quantum kernels places our proposed method  tractable even for a large number of qubits.

\section{The PQAE and QAE models \label{S3}}

In this section we present  PQAE and QAE models. The first is a pre-processing method of classical data
which is adjusted to the data-embedding part of a VQC. This has a hybrid classical-quantum structure and
 requires access to the specialized hardware of the quantum circuit. QAE consists of  a feature map emulating a quantum feature map, embedded in the bottleneck layer of a classical AE. QAE  runs exclusively on classical hardware. 

\subsection{PQAE for encoding on qubit states}

The goal of PQAE is to transform classical data into a format that quantum circuits can  efficiently use. The PQAE learns how to compress and encode classical data into this format by passing it through the feature map and then reconstructing it back to the original dataset. To do this effectively, it uses the same feature map as the VQC that will process the data later. 
 
\begin{figure*}[t]
    \centering
    \hspace*{-0.15\textwidth}
    \includegraphics[ width=1.3\textwidth, height=0.35\textheight]{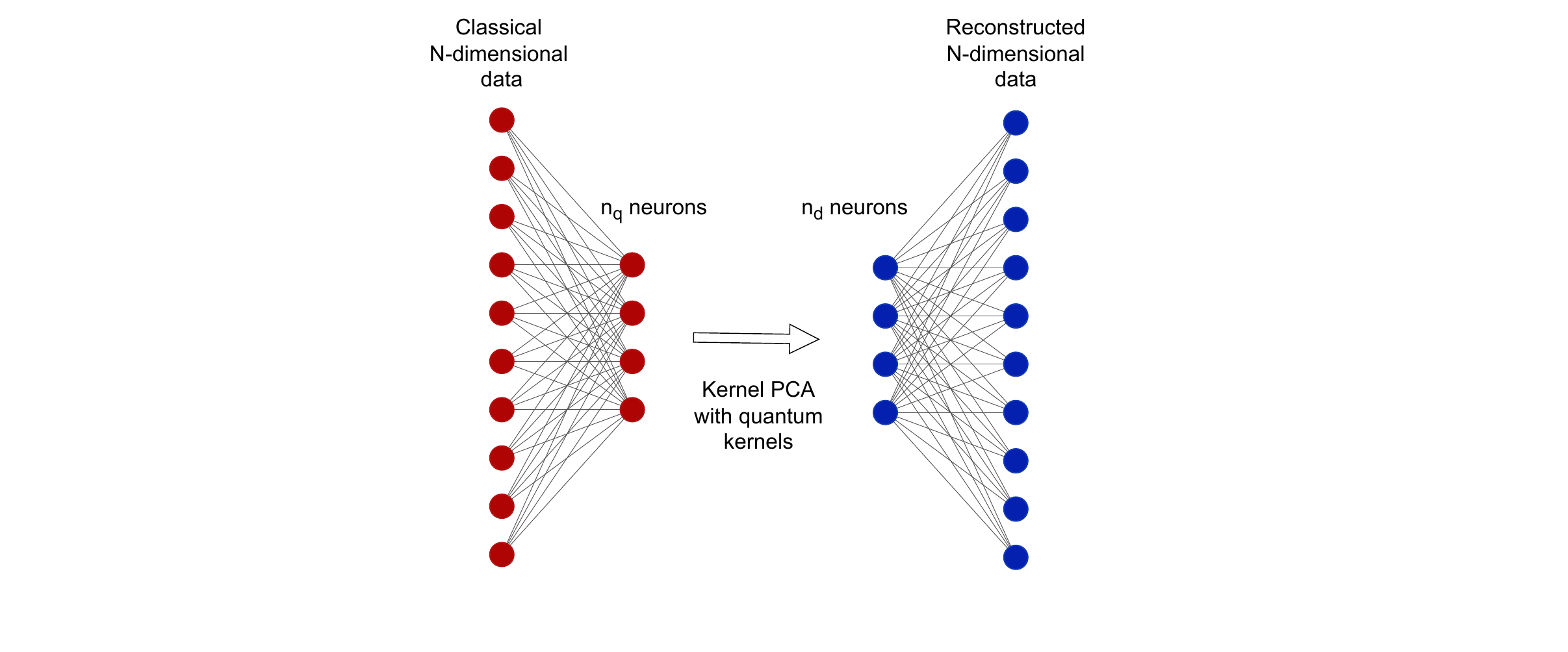}
    \caption{PQAE: Classical $N$-dimensional data are encoded to the classical nodes on the left of the network. The dimensionality is reduced to $n_q$ with classical means and loaded on $n_q$ qubits to be processed by a data-encoding layer. The output is used to construct quantum kernels which are then used for a classical KPCA.  Finally, the  $n_d$ principal components   are   passed through classical neurons to reproduce the original dimensionality of data. During the training the parameters in the classical network are
    optimized so that MSE between input and output data is minimized}
    \label{fig3}
\end{figure*}

Given the data-encoding block of a VQC of $n_q$ qubits and an $N$-dimensional train dataset,  PQAE is an algorithmic procedure that employs a classical network and a quantum circuit, see Fig.~\ref{fig3}. The goal of the discussed machine learning method is to adjust the input parameters of the classical network, so that the overall loss -- in this case given by the MSE -- is minimised across the dataset. The procedure comprises (iteratively) repeated epochs and this stops when convergence in the value of MSE is achieved. For each epoch the train dataset is randomly divided in batches of $M$ data, and for each batch the updating of the classical weights is performed as follows:  
\begin{itemize}
    \item  Every data point of the batch is passed via the same classical linear layers, which have the structure of the encoding layer of an AE, see Fig.~\ref{fig3}.  The aim of this step is to adjust the dimensionality of data, $N$, to the number of qubits,  $n_q$.
    \item The values on the $n_q$ neurons are then used as features for the data-encoding layer of the VQC under study.
     Each $i$ data  point produces a quantum state $|\psi_i\rangle$,  with $i=1,\ldots M$.
    \item  A SWAP or Inversion test is used to estimate the $M \times M$ quantum kernel matrix $k_{i,j}= |\langle \psi_i | \psi_j \rangle|^2$.
    \item  Afterwards KPCA is performed using the quantum kernel $k_{i,j}$ and this outputs $n_d \leq M$  principal components of dimension $M$.
    \item The elements of principal components are attributed to the related data of the batch and one proceeds by passing this $n_d$-dimensional vector via  a series of classical hidden layers   until the original dimensionality $N$ is restored.
    \item The  mean squared error (MSE) is calculated for the batch, Eq.(\ref{eq1}), and the weights in classical encoding and decoding layers are updated according to a gradient optimization method.
\end{itemize}
One  repeats the procedure for all batches and then proceed to the next epoch.

We note here that practically the number of qubits $n_q$ can be close to  the number of features $N$, when $N$ is small, or when technological advances allow for more qubits. Moreover, the latent dimension $n_d$ can be considered greater than $n_q$ the emdedding is performed in
in the Hilbert space  of dimension $2^{n_q}$. Therefore, without loss of generality we assume  that $n_q \le n_d < N$. Finally in this work  we  restrict the structure of classical encoding and decoding layers to linear ones.

Let us describe a case example in order to further clarify the  PQAE algorithmic procedure. Suppose that a dataset with $N = 100$ features is given and they want to use a VQC with $n_q = 4$ qubits and a ZZ feature map. The ZZ feature map was introduced in \cite{QAE4} and has been used successfully in \cite{QAE33,QAE34} for solving classification tasks . In order to efficiently encode these 100 features into 4 qubits the PQAE method is  used. They start with a hidden layer that maps 100 nodes to 32 nodes, and then another hidden layer that maps 32 nodes to 4 nodes. This could be alternatively done   through multiple hidden layers or directly from 100 to 4. The 4 nodes are then passed to a quantum computer using the ZZFeatureMap for encoding. The output data are now encoded  on   qubits,  and they  calculate the kernel matrix using the quantum algorithmic procedure of SWAP or Inversion test. They then perform KPCA  with the quantum kernels to reduce the data in the transformed space. Let one select the top $4$ principal components ($n_d=4$).  Finally, they use linear layers to decode from $n_d=4$ to 32, and then from 32 to 100, reproducing the input data. The weights of the connections between nodes are trained to reduce MSE. After training, the PQAE model can be used to encode new unseen test data into  VQC, effectively serving as a ``qubit encoding'' black box.

\subsection{QAEs for dimensionality reduction \label{QAE}}

In the PQAE method the KPCA is used to extract information  from a large Hilbert space and bring it back to
the classical space in a compact way. Thus this component is necessary for the accessibility of the method in the case
of many qubits. Here, we  re-consider the model omitting the KPCA. Unavoidably the model
that we build is not useful for encoding in many qubits, but on the other hand this can be seen as an unconventional classical AE 
where its bottleneck layer is `dressed' with a quantum feature map. In the next section, we investigate  its advantages numerically.

\begin{figure*}[t]
    \centering
    \hspace*{-0.25\textwidth}
    \includegraphics[ width=1.5\textwidth, height=0.35\textheight]{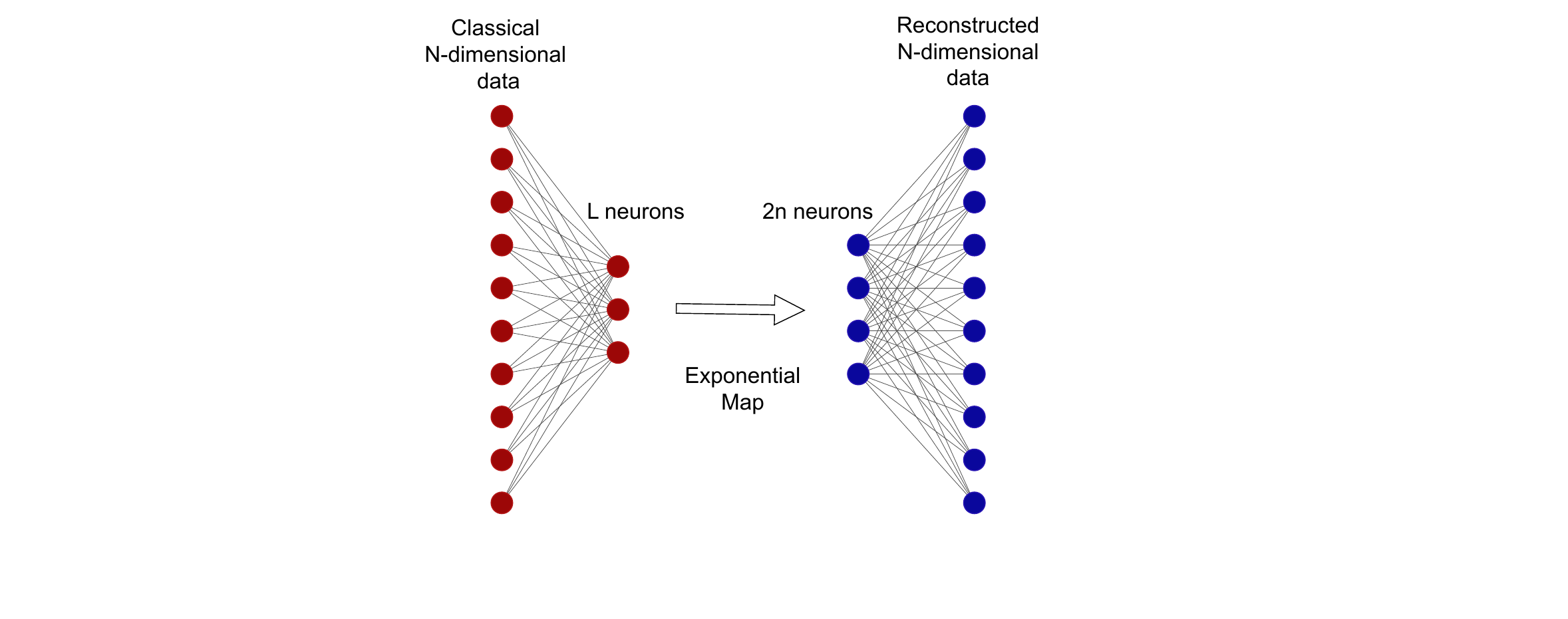}
     \caption{QAE: The network starts with $N$ classical nodes and reduces to  $L=n^2 -1$ dimensions. The $L$ dimensional data are used to parametrize $SU(n)$ rotations and create an $n\times n$ unitary matrix. One extracts $2n$ real numbers from this matrix which are used to parametrize $2n$ neurons. Then this data is decoded via a layered network back to the original dimensions.
    }
    \label{fig4}
\end{figure*}

QAEs begin by reducing the dimensionality of the input $N$-dimensional data to a smaller set of $L$ neurons. This reduction step is analogous to the initial compression phase in classical AEs and the encoded neurons are represented by Eq.(\ref{eq2}). The operator that creates the next layer of neurons, which is derived from the output of the feature map, is represented with Eq.(\ref{eq6}) \begin{equation}e^{-i \sum_{j=1}^{L} z_j \hat{g}_j } ~.\label{eq6}
\end{equation} where the variables, $z_j$ with $j=1,\ldots,L$ are used to parametrize an $SU(n)$ rotation and the $\hat{g}_j$ operators are a subset of the $n^2 - 1$ generators of the related Lie algebra \cite{QAE23}. For simplicity let us set here $L=n^2-1$ and the application of the method to the more general case $L<n^2-1$ is, then, straightforward. The features of the initial data points are encoded in a unitary matrix of the form

\begin{equation}
\begin{pmatrix}
a & b \\
-b^* & a^*
\end{pmatrix}\label{eq7}
\end{equation}    
with \( a \),\( b \) $\in \mathbb{C}$ satisfying $|a|^2+|b|^2=1$, which is the $SU(2)$ case (or single qubit case).  We extract the elements of the first column of this $n\times n$ matrix, which can is equivalent to the action of operating  the  unitary matrix on the ground state of an $n$-dimensional quantum system.
Alternatively one can use the average values over each row's elements, that corresponds to the operation on a state that is in equal superposition over all states of the computational basis. Either way, one obtains $2n$ real numbers, which for $n>2$ are parametrically dependent to each other
and adhere to the normalization condition. These real numbers are encoded on $2n$ neurons and then the decoding starts until the original dimensionality is reached. We note that, the dimensionality of the bottleneck of any autoencoder, is determined by the layer with the fewest neurons. Only in the $SU(2)$ case, this occurs in the layer before the feature map. In all other cases, where $n>2$, the bottleneck is in the layer after the feature map. The whole procedure is schematically presented in Fig.~\ref{fig4}.

Since QAEs aim to offer a fruitful representation in the latent space, batched QAEs form a variant where input data is processed in $m$ batches, enhancing scalability and performance. Following Fig.~\ref{fig5}, segments of the reduced representation of the data is passed through the feature map. For example, starting with an $100$  dimensional input, one can reduce the dimension to $12$ with classical layers and separate these features in $m=4$ batches. Then, using four Pauli feature maps with 3 generators per batch, 16 real outputs are produced and processed for decoding via the classical layers. In this way, the feature map is essentially used more than once, offering greater encoding capabilities. The non-batched case presented earlier is a special case of batched QAEs with $m=1$.

\begin{figure*}[t]
    \centering
    \hspace*{-0.3\textwidth}
    \includegraphics[ width=1.55\textwidth, height=0.35\textheight]{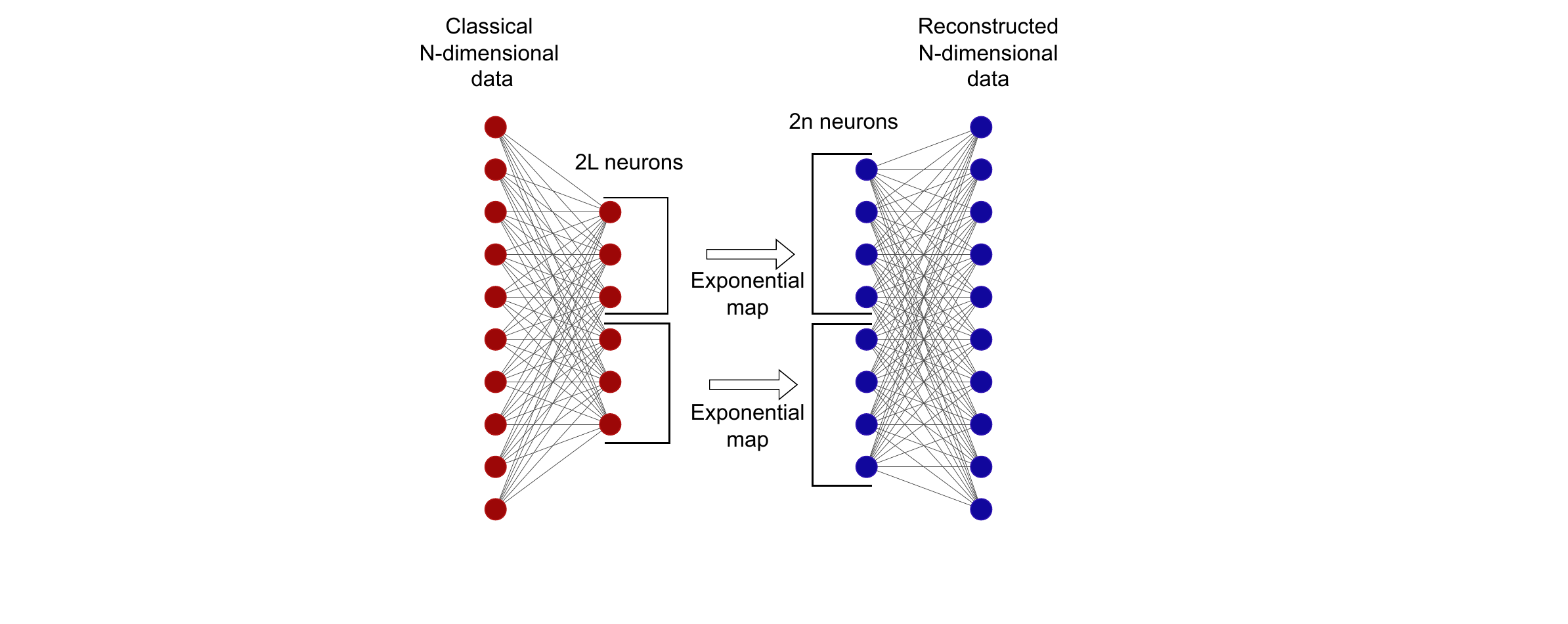}
    \caption{Batched QAE: The exponential map is performed in two subsets ($m=2$) of the reduced data. }
    \label{fig5}
\end{figure*}

\section{Numerical Results \label{S4}}
In order to exhibit the efficacy of the proposed methods we first apply the PQAE model for encoding classical data on a VQC and  compare the results with those obtained via PCA.  We then apply QAE methodology for dimensionality
reduction of data and we compare its reconstruction capability with  the one of classical linear AEs. For all tasks, PyTorch \cite{QAE36} is used to construct the AEs and the ADAM and RMSprop optimizers are used for training the PQAE and QAE models accordingly. The VQCs for the Iris, Wines and Seed dataset, are implemented in IBM's Qiskit \cite{QAE37}. All programs and datasets are available on GitHub \cite{QAE35}.

\subsection{Encoding  VQCs with PQAE method }

VQCs in IBM's Qiskit are composed of a Pauli feature map in the data-encoding block and a parametrized ansatz in the variational block. In this approach, we use a PQAE to encode the classical data into qubits for four different datasets available in scikit-learn \cite{QAE38}, namely Iris  ($N=4$), Wines ($N=13$), Seed ($N=7$) and MNIST ($N=64$). The transformed data are then passed to a VQC in order to solve the corresponding classification task. For all datasets, we set-up a $n_q=4$ qubit VQC using Qiskit's ZZFeatureMap with 2 repetitions and a Real Amplitudes ansatz with 3 repetitions. The classical encoding part of the PQAE consists of one linear layer which transforms the data from $\mathbb{R}^N$ to $\mathbb{R}^{n_q}$ and the decoding part consists of a linear layer which transforms the reduced space $\mathbb{R}^{n_d}$ to $\mathbb{R}^N$.  The optimization procedure is repeated using 10 different random seeds and we collect the results with  best accuracy for both PQAE and PCA methods. In both cases, we train the VQC for $100$ epochs. The results are reported in Table~\ref{tab:mytable1}. PQAEs increased the VQC's accuracy across three out of four datasets, as compared to PCA or Raw encoding, with the most remarkable achievement in accuracy being 97\% in the Iris dataset, even surpassing the accuracy of qutrit-based quantum neural networks, which aspire to offer an improvement over qubit methods \cite{QAE43}.

\begin{table}
    \centering
    \begin{tabular}{|c|c|c|c|c|}
        \hline
        & Iris$^*$ & Wines & Seed & MNIST \\
        \hline
        PCA or Raw$^*$ encoding & 0.77 &  0.56 &  0.57 &  0.82\\
        \hline
        PQAE encoding & 0.97 &  0.83 & 0.74 &  0.81\\
        \hline
       
    \end{tabular}
    \vspace{0.5cm}
    \caption{Classification accuracy achieved via VQC  for different datasets. To encode  data on VQC we use  PCA (or raw endoding in the case of Iris set) and PQAE encoding.}
    \label{tab:mytable1}
\end{table}

The advantage exhibited by PQAE stems from the fact that this pre-processing model is trained to minimize the reconstruction MSE, ensuring that the encoded data retain the essential features of the original data while filtering out multicollinearity and irrelevant information. This training acts as an unsupervised feature extraction mechanism, resulting in a cleaner and more manageable data representation. Furthermore, by utilizing the same quantum feature map in both the PQAE and the VQC ensures consistent data transformation, aligning the features learned during the autoencoder's training with those used by the VQC. This consistency allows the VQC to focus on fine-tuning its decision boundaries rather than representing the data itself. Consequently, the dimensionality reduction achieved by PQAE enables the VQC to operate on a more compact and informative data representation, leading to faster convergence and better generalization. 

The combination of classical and quantum methods in the approach that we present  leverages the strengths of both. 
The classical autoencoder excels at unsupervised feature extraction and multicollinearity reduction, while the quantum feature map and VQC capitalize on the expressive power of quantum computations for classification.

\subsection{QAEs vs classical linear AE for dimensionality reduction}

QAEs provide an exploration primitive of the synergy between quantum and classical ML techniques. By integrating quantum feature maps into classical autoencoder architectures, we investigate the improvement prospects of data processing tasks based on quantum mechanics over classical  alternatives, while providing numerical results. The comparison of reconstruction error between QAE and AEs provides a framework for analyzing the expressive power of given quantum transformations.

\begin{figure*}[t]
    \centering
    \hspace*{-0.3\textwidth}
    \includegraphics[ width=1.65\textwidth, height=0.45\textheight]{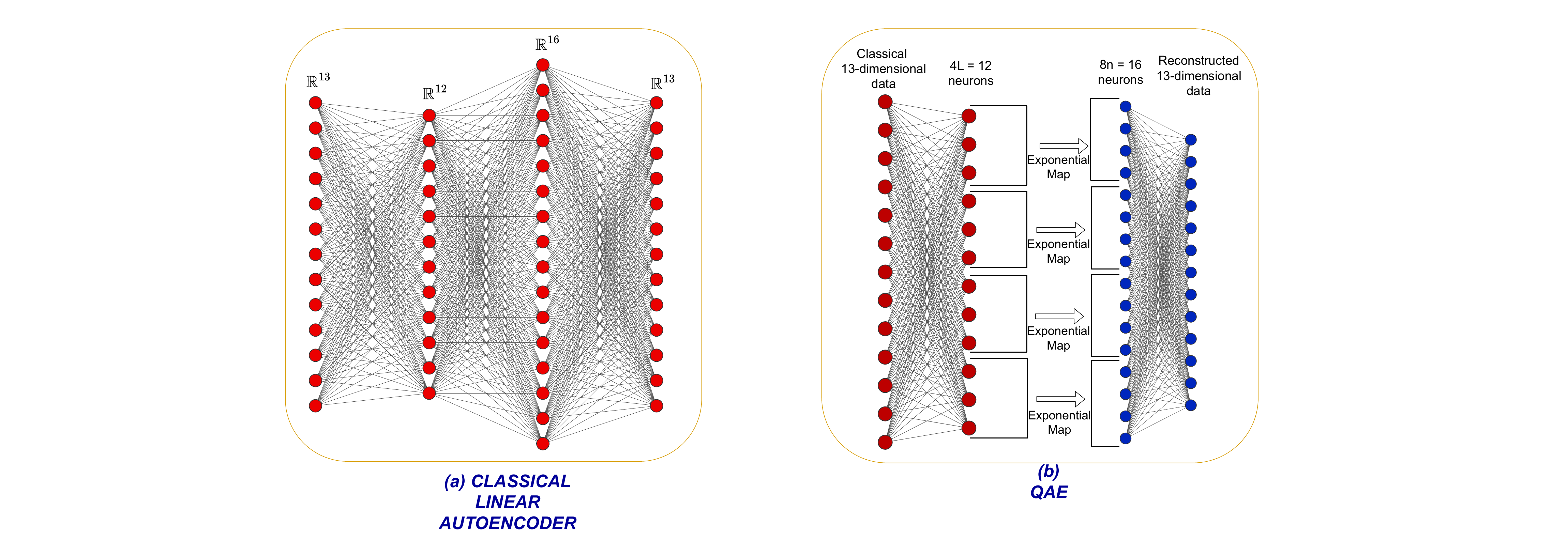}
    \caption{Wines dataset is treated via two models: (a) a classical linear AE and (b) a batched QAE. The reduction of the feature space in both cases is from $\mathbb{R}^{13}$ to $\mathbb{R}^{12}$ on the bottleneck. The insertion of the $\mathbb{R}^{16}$ layer on the linear AE cannot change the achieved MSE and this has been added so that the QAE's and AE's architectures are in direct analogy.}
    \label{fig6}
\end{figure*}

The output of the exponential map used in QAE produces features which follow an $l^2$-normalization, and thus it is sensible to apply the method on data having this same property. For specific types of data, e.g. wavefunctions, this property  naturally occurs, for others this can be imposed by taking the extra step of normalization. All models in this study are compared with classical linear AEs which follow the exact same architecture as the QAEs, except for the feature map, see for instance Fig.~\ref{fig6}, subfig.~(a).

We start our numerical investigations with the  Iris dataset, that is first normalized. We use QAE
to reduce its feature space from $\mathbb{R}^4$ to $\mathbb{R}^3$ at the bottleneck, and in Table~\ref{tab:mytable3} we compare the results, i.e, MSE, with the one achieved by a classical AE of  same architecture. The Seed dataset is treated in a similar way, and its feature space is reduced  from $\mathbb{R}^{7}$ to $\mathbb{R}^{3}$.
In Table~\ref{tab:mytable3} we exhibit that QAE can offer improvement to non normalized data, as well, using  the Wines dataset, with the dimensionality of data reduced from $\mathbb{R}^{13}$ to $\mathbb{R}^{12}$. We achieve this result by using a batched QAE with $m = 4$ batches. As depicted in Fig.~\ref{fig6}, subfig.~(b), QAE starts with a `classical' layer which reduces the dimensionality from $\mathbb{R}^{13}$ to $\mathbb{R}^{12}$. This  is followed up by  four feature maps. The output of the feature maps provides the values for a layer of $16$ neurons. A decoding layer follows reducing the dimensions  back to $\mathbb{R}^{13}$. This procedure  achieves a reduction greater than 50\% on the MSE  achieved by  the corresponding classical AE of Fig.~\ref{fig6}, subfig.~(a).

Furthermore, we implement a QAE model with an alternative feature map to the one presented in Section~\ref{QAE}. In the latter case, the feature map described by the Eqs.(\ref{eq6})-(\ref{eq7}) induces an encoding from \( \mathbb{R}^3 \) to \( \mathbb{R}^4 \). Here, we experiment with a quantum-inspired map that we name Bloch encoding, which gets a 2-dimensional input and outputs a 3-dimensional vector, i.e., 
 \( \mathbb{R}^2\rightarrow\mathbb{R}^3 \). This is achieved via the Bloch vector representation of a qubit: \begin{equation}
     |\psi\rangle = \cos\left(\frac{\theta}{2}\right) |0\rangle + e^{i\phi}\sin\left(\frac{\theta}{2}\right) |1\rangle \end{equation} 
 where as input we consider the angles $\theta$, $\phi$ and as  output
 the three real numbers: $\left\{\cos\left(\frac{\theta}{2}\right),~ \sin\left(\frac{\theta}{2}\right) \cos\left(\phi\right),~\sin\left(\frac{\theta}{2}\right) \sin\left(\phi\right) \right\}$.
 The latent space offered by such a feature map is thus further reduced from $3$ to $2$. 
We apply the  QAE model equipped with the Bloch feature map to the normalized Iris dataset so as to reduce the bottleneck dimension to 2. The improvements over AE can be found in Table~\ref{tab:mytable4} where one may see that this type of QAE bottleneck reduces the MSE  by almost $2.5$ times.

\begin{figure*}[t]
    \centering
    \hspace*{-0.1\textwidth}
    \includegraphics[ width=0.9\textwidth, height=0.35\textheight]{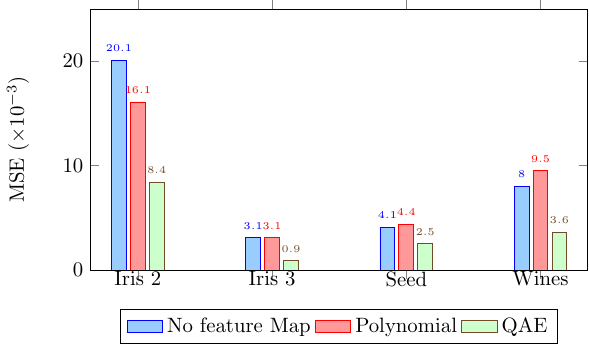}
    \caption{Comparing MSE reached by linear, polynomial and quantum inspired feature maps. We investigate the Iris, Seed  and the  Wines dataset (with $m=4$ batches).
Iris dataset is treated twice: Iris 2 refers to the Bloch QAE model (Table~\ref{tab:mytable4}) of latent dimension 2 and Iris 3 refers to the  QAE model (Table~\ref{tab:mytable3}) of latent dimension 3.}
    \label{fig7}
\end{figure*}

Finally in Fig.~\ref{fig7} we compare the QAEs' results  presented in Tables~\ref{tab:mytable3}-\ref{tab:mytable4}  with those  reached by AEs equipped with polynomial feature map. Notably, the latter  exhibit very good  performance as compared to other kernels \cite{QAE41,QAE42}. In more details we use the map $\left\{x_{1},~x_{2}\right\}\rightarrow\left\{x_{1}^{2},~x_{2}^{2},~x_{1}x_{2}\right\}$ for the \( \mathbb{R}^2\rightarrow\mathbb{R}^3 \) case and the slightly altered version of the habitual polynomial feature map $\left\{x_{1},~x_{2},~x_{3}\right\}\rightarrow\left\{x_{1}^{2},~x_{2}^{2},~x_{3}^{2},~x_{1}x_{2}+x_{1}x_{3}+x_{2}x_{3}\right\}$, to match the \( \mathbb{R}^3\rightarrow\mathbb{R}^4 \) case. In every dataset, the comparisons show an advantage for the QAE method over the polynomial feature maps.

\begin{table}
    \centering
    \begin{tabular}{|c|c|c|c|c|}
        \hline
        & Iris MSE ($l^2$) & Seed MSE ($l^2$) & Wines MSE (batched) \\
        \hline
       Linear AE & 0.0031 & 0.0041 &  0.0080\\
        \hline
        QAE  & 0.0009  & 0.0025 &  0.0036 \\
        \hline
        Latent Dimension  & 3  & 3 &  12 \\
        \hline
       
    \end{tabular}
    \caption{Data reproduction using QAE method as compared to linear AE methods. The reduced latent dimension is the same
    in both methods.}
    \label{tab:mytable3}
\end{table}

\begin{table}
    \centering
    \begin{tabular}{|c|c|c|}
        \hline
        & Iris MSE (latent=2)   \\
        \hline
       Linear AE & 0.0201  \\
        \hline
        QAE  & 0.0084   \\
        \hline
       
    \end{tabular}
    \caption{ QAE with Bloch encoding for the Iris dataset.}
    \label{tab:mytable4}
\end{table}

\section{Discussion \label{S5}}
In this work we have introduced PQAE, a method for reducing the dimensions of features of classical data sets that is tailored to the
quantum feature map of a VQC. The method uses a hybrid structure: a classical linear AE with the bottleneck layer made of  the quantum feature map
of the data-embedding circuit of the VQC.
The input of the quantum feature map is a batch of train data which have passed through the encoding layer of the AE.  Its output is the quantum kernel matrix for the batch  that is consequently  used for KPCA. The vector comprising the principal components of the quantum kernel matrix
is the output treated by the decoding layer of the AE. 
The numerical results of  Section~\ref{S4}
support the claim that this hybrid method  of pre-processing can considerably boost the performance of a VQC in solving classification tasks.
In addition, the quantum-inspired model  QAEs and batched QAEs,    create new possibilities for dimensionality reduction of classical data, since we have identified cases where they  greatly outperform simple linear AEs of the same latent dimension.

There are several points which can be further investigated in order to potentially  improve the performance of the proposed models. In this work, we employ MSE as a measure of reproducibility of datasets but other  measures could be proven more fit to our methods. We also exclusively use a linear structure for the classical layers in the models as well as for the classical AEs with which we compare the results.   An extension to nonlinear activation functions seems necessary for completeness. Furthermore, all numerical examples in this work concern quantum circuits made of qubits even though the methodology supports an extension to qudits. In a future work where more
complicated data sets and multi-class problems are investigated, we would be interested in employing the methods for qudit circuits. Finally, PQAE has been developed and exhibited for variational quantum classifiers but the methodology could be adapted to other variational algorithmic procedures, such as variational quantum regressors, which also require uploading of classical data on quantum circuits.

\section*{Acknowledgements}
This work was supported by   the project Hellas QCI co-funded by the European Union under the Digital Europe Programme  grant agreement  No.101091504. A.M. and D.S. acknowledge partial support from the European Union’s Horizon
Europe research and innovation program under grant agreement No.101092766 (ALLEGRO Project).

\bibliography{references}

\end{document}